
\magnification=\magstep1
\font\logo=logo10 
\font\logosl=logosl10 
\font\sc=cmcsc10 

\def\MF{{\logo META}\-{\logo FONT}}
\def\MFbook{{\sl The {\logosl METAFONT}\kern1pt book}}
\def\adj{\mathrel-\joinrel\mathrel-}   
\def\disleft#1:#2:#3\par{\par\hangindent#1\noindent
			 \hbox to #1{#2 \hfill \hskip .1em}\ignorespaces#3\par}

\input picmac
\newcount\ii \newcount\jj
\unitlength=10pt
\def\dsk{\disk{.3}}
\def\upvec(#1,#2){\put(#1,#2){\dsk} \put(#1,#2){\vector(0,1)1}}
\def\lfvec(#1,#2){\put(#1,#2){\dsk} \put(#1,#2){\vector(-1,0)1}}
\def\dnvec(#1,#2){\put(#1,#2){\dsk} \put(#1,#2){\vector(0,-1)1}}
\def\rtvec(#1,#2){\put(#1,#2){\dsk} \put(#1,#2){\vector(1,0)1}}

\parskip 2pt
\vskip-3\baselineskip

\centerline{\bf The Problem of Compatible Representatives}
\centerline{by Donald E. Knuth\footnote{*}{Computer Science
Department, Stanford University; research supported in part by
National Science Foundation grant CCR--8610181.}
and Arvind Raghunathan\footnote{**}{University of California, 
Davis, Division of Computer Science; research supported in part by 
Semiconductor Research Corporation grant SRC--82--11--008.}}

\bigskip
\bigskip
{\narrower\smallskip\noindent
{\bf Abstract.}\enspace
The purpose of this note is to attach a name to a natural class of
combinatorial problems and to point out that this class includes many
important special cases. We also show that a simple problem of placing
nonoverlapping labels on a rectangular map is  NP-complete.
\smallskip}

\vfill

(Combinatorial programming)

\bigskip
\line{{\bf Abbreviated title}\enspace Compatible Representatives\hfil}

\eject

\centerline{\bf The Problem of Compatible Representatives}
\centerline{by Donald E. Knuth
and Arvind Raghunathan}

\bigskip
Many combinatorial tasks can be formulated in the following
way: Is there a sequence $(x1,x2,\ldots,xn)$ such that $xj\in Aj$
for all~$j$, and~$xj$ is compatible with~$xk$ for all $j<k$? Here
$A1,A2,\ldots,An$ are given sets, and ``compatibility'' is a given
relation on $A1A2\ldots An$.

This problem is NP-hard in general. For example, if all sets $Aj$ are the
same, and if compatibility is a symmetric, irreflexive relation, 
a~sequence of compatible representatives is nothing but an $n$-clique
in the compatibility graph. 

The problem of coloring a graph $G$ with $c$ colors is another NP-hard special
case of the general compatibility question. Let $Aj$ be the the set of
pairs $\{(j,1),\ldots,(j,c)\}$, and say that $(j,a)$ is compatible with
$(k,b)$ if either $ab$ or $vj$ is not adjacent to~$vk$ in~$G$, where the
vertices of~$G$ are $\{v1,\ldots,vn\}$. Then a sequence of compatible
representatives is essentially a $c$-coloring of~$G$.
Therefore the problem is NP-hard for all $c3$.

On the other hand, the compatibility problem also has important special
cases that are efficiently solvable. 
If the compatibility relation is `$\neq$', then a solution sequence
$(x_1,\ldots,x_n)$ is traditionally called a system of distinct
representatives [4] [3, Chapter~5], and the problem of finding such
systems is well known to be equivalent to bipartite matching. Indeed, 
if the compatibility relation is the
complement of any equivalence relation, a~sequence $(x1,x2,\ldots,xn)$
of compatible representatives exists if and only if there is a matching of
cardinality~$n$ in a bipartite graph on the vertices
$\{v1,\ldots,vn,c1,\ldots,cm\}$, where $\{c1,\ldots,cm\}$ are the
equivalence classes and we have $vj\adj ck$ if and only if~$Aj$
contains an element of class~$ck$.

Another nice special case is equivalent to identifying increasing
subsequences of a permutation. Let $\pi1 \ldots \pim$ be
a permutation of $\{1,\ldots,m\}$, and let $Aj$ be the set of pairs
$\{(j,1),\ldots,(j,m)\}$. Say that $(j,a)$ is compatible with $(k,b)$
if and only if $j<k$ and $\pia<\pib$.
Then a compatible sequence $\bigl((1,a1),\ldots,(n,an)\bigr)$ is
equivalent to an increasing subsequence $(\pi{a1},\ldots,\pi{an})$
of $\pi1\ldots \pim$.

The example in the prevous paragraph illustrates that compatibility need
not be a symmetric relation. But when the sets~$Aj$ are pairwise disjoint,
as in that case, we could just as well assume that compatibility is
symmetric and reflexive, since our definition of compatible representatives
makes it immaterial whether elements~$xj$ of~$Aj$ and $xk$ of~$Ak$ are
compatible unless we have $j<k$.

There are, however, important special cases in which compatibility is
asymmetric. Consider, for example, a~scheduling problem in which $Aj$
is a set of tasks that can be done at time~$j$, and where $xj$ is
compatible with~$xk$ only when task~$xj$ does not require the prior
completion of~$xk$.

Cartographers face an interesting case of the general compatibility
problem when they attach alphabetic labels to dots on a map. Let $Aj$
represent the possible ways to place the name of city~$j$, and let $xj$
be compatible with~$xk$ when positions~$xj$ and~$xk$ do not overlap
each other or otherwise mislead a potential reader. Then a good map should be
a solution to the problem of compatible representatives.

Notice that the cartographic problem makes sense even if the sets $Aj$
are infinite. The task 
of placing disjoint labels
is a fairly natural question of combinatorial geometry
that does not appear to be a special case of any other well known
problem.

In light of this discussion,
 it seems worthwhile to add the problem of compatible
representatives to the class of ``combinatorial problems that deserve
a name,'' and to investigate heuristics and additional special cases that
turn out to have efficient solutions.

\bigskip\noindent
{\bf Simple special cases.}\enspace
We have noted that the compatibility problem is equivalent to bipartite
matching when incompatibility is an equivalence relation. The problem also
has a polynomial-time solution when compatibility is transitive. Let
$B1=A1$, and for $j>1$ let 
$$Bj=\{\,y\in Aj\mid \exists\,x\in B{j-1}\;(x\hbox{ compatible with $y$})
\,\}\,.$$
Then the compatibility problem has a solution if and only if $Bn$ is
nonempty. We can decide this in at most $\sum{j=2}^n\|A{j-1}\|\,\|Aj\|$ steps.

Another noteworthy special case occurs when each set $Aj$ contains at most
two elements. Then the compatibility problem is equivalent to an instance
of {\tt 2SAT}: We can assume that $Aj=\{vj,\overline{v}j\}$;
 the clauses
are $(\overline{\sigma}j\vee \overline{\sigma}k)$ for every pair of
literals such that $j<k$ and $\sigmaj$ is incompatible with~$\sigmak$.

In general, if each $\Vert A_j\Vert\leq k$ and $k\geq 2$, the problem
reduces directly to an instance of {\sl k\/}{\tt SAT} in which each
literal occurs positively just once. The literals are $(j,a)$ for
$a\in A_j$, and the clauses are
$$\vcenter{\halign{$\hfil{\displaystyle{#}}$\qquad&#\hfil\cr
\bigvee_{n\in A_j}(j,a)\,,&for $1\leq j\leq n\,;$\cr
\noalign{\smallskip}
\overline{(j,a)}\,\vee\,\overline{(k,b)}\,,&for $1\leq j<k\leq n$ and
$a$ incompatible with $b$.\cr}}$$  

Conversely, any instance of {\sl k\/}{\tt SAT} with $m$ clauses
reduces to the compatibility problem of finding representatives
$(x_1,\ldots,x_m)$, with $x_j$ a member of the $j\/$th clause and with
two literals compatible iff they aren't negatives of each other.

The general compatibility problem with finite sets~$A_j$ can also
 be reduced to an independent set problem
in a natural way. Consider the graph~$G$ with vertices $(j,a)$ for
$a\in Aj$, having edges
$$\eqalign{&(j,a)\adj(j,b)\,,\qquad\hbox{if $a\neq b$};\cr
&(j,a)\adj(k,b)\,,\qquad\hbox{if $j<k$ and $a$ is incompatible with $b$}.\cr}$$
Then $G$ has an independent set of size $n$ if and only if the compatibility
problem has a solution.

Therefore we obtain simple solutions of the compatibility problem when there is
a simple solution to the corresponding independent set problem. One such case
occurs when compatibility is a symmetric relation and we have the
following condition:
If $i<j<k$ and the elements
$a_i,a_j,a_k$ are mutually compatible, then (1)~every element of~$Ai$ is
compatible with either~$a_j$ or~$a_k$; (2)~every element of~$A_j$ is
compatible with either $a_i$ or~$a_k$; (3)~every element of~$A_k$ is 
compatible with either~$a_i$ or~$a-j$; and (4)~every element not in 
$A_i\cup A_j\cup A_k$ is compatible with either $x_i$, $x_j$, or~$x_k$.
In such a case the  graph~$G$ is claw-free, and we can use Minty's
algorithm~[7] to find a maximum independent set.

Gr\"otschel, Lov\'asz, and Schrijver [2, Chapter 9] have compiled a survey
of cases where the independent set problem is known to have a simple solution.

\bigskip\noindent
{\bf Another hard case.}\enspace
A very special case of the general mapmaker's problem, alluded to in the
introduction, turns out to be NP-complete.

Consider a set of integer points $p1,\ldots,pn$ on the plane. We wish to
find integer points $x1,\ldots,xn$ with the following properties for
all $jk$:
$$|xj-pj|=1\,;\qquad |xj-pk|>1\,;\qquad |xj-xk|2\,.$$
(Motivation: Each $xj$ is the center of a $2\times 2$ square in which 
a ``label'' for point~$pj$ can be placed. The label at~$xj$ 
should be closer
to~$pj$ than to any other point; distinct labels should not overlap.) We will
call this the {\tt MFL} problem, for ``\MF\ labeling,'' because it arises
in connection with the task of attaching labels to points in diagrams
drawn by \MF\ [5, page 328].

Solutions to the {\tt MFL} problem can conveniently be represented by
showing each point~$pj$ as a heavy dot and drawing an arrow from~$pj$
to~$xj$ for each~$j$; at most four possibilities exist from each of the
given points. For example, it is easy to see that a cluster of four
adjacent points can be labeled in only two ways:
$$
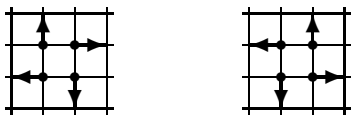
\beginpicture(3.5,3.5)(-.25,-.25) \thinlines
\multiput(-.2,0)(0,1)4{\line(1,0){3.4}}
\multiput(0,-.2)(1,0)4{\line(0,1){3.4}}
\thicklines
\lfvec(1,1) \upvec(1,2) \rtvec(2,2) \dnvec(2,1)
\endpicture
\hskip 4\unitlength
\beginpicture(3.5,3.5)(-.25,-.25) \thinlines
\multiput(-.2,0)(0,1)4{\line(1,0){3.4}}
\multiput(0,-.2)(1,0)4{\line(0,1){3.4}}
\thicklines
\lfvec(1,2) \upvec(2,2) \rtvec(2,1) \dnvec(1,1)
\endpicture
$$
There is no way to attach a label to the middle point in a configuration like
$$
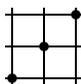
\beginpicture(2.5,2.5)(-.25,-.25) 
\multiput(-.2,0)(0,1)3{\line(1,0){2.4}}
\multiput(0,-.2)(1,0)3{\line(0,1){2.4}}
\put(0,0){\dsk}
\put(1,1){\dsk}
\put(2,2){\dsk}
\endpicture
$$
because each of the four positions adjacent to that point is too close to
one of the other given points. The {\tt MFL} problem provides an amusing
pastime for people who are sitting in a boring meeting and who happen to have a
tablet of graph paper to doodle~on.

The general {\tt MFL} problem is clearly in NP\null.
In order to show that it is NP-complete, we observe first that there are only
two solutions to the problem
$$
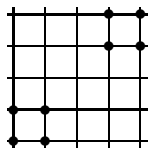
\beginpicture(4.5,4.5)(-.25,-.25)
\multiput(-.2,0)(0,1)5{\line(1,0){4.4}}
\multiput(0,-.2)(1,0)5{\line(0,1){4.4}}
\put(0,0){\dsk}
\put(1,1){\dsk}
\put(0,1){\dsk}
\put(1,0){\dsk}
\put(3,3){\dsk}
\put(3,4){\dsk}
\put(4,3){\dsk}
\put(4,4){\dsk}
\endpicture
$$
namely the two solutions for four-point clusters given earlier, using the
same orientation in each cluster. Thus we can construct large chainlike
tree networks of four-point clusters, for example,
$$\unitlength=8pt
\def\dsk{\disk{.4}}
\def\\(#1,#2){\ii=#1 \jj=#2
 \put(\number\ii,\number\jj){\dsk}
 \advance\ii1 \put(\number\ii,\number\jj){\dsk}
 \advance\jj1 \put(\number\ii,\number\jj){\dsk}
 \advance\ii-1 \put(\number\ii,\number\jj){\dsk}}
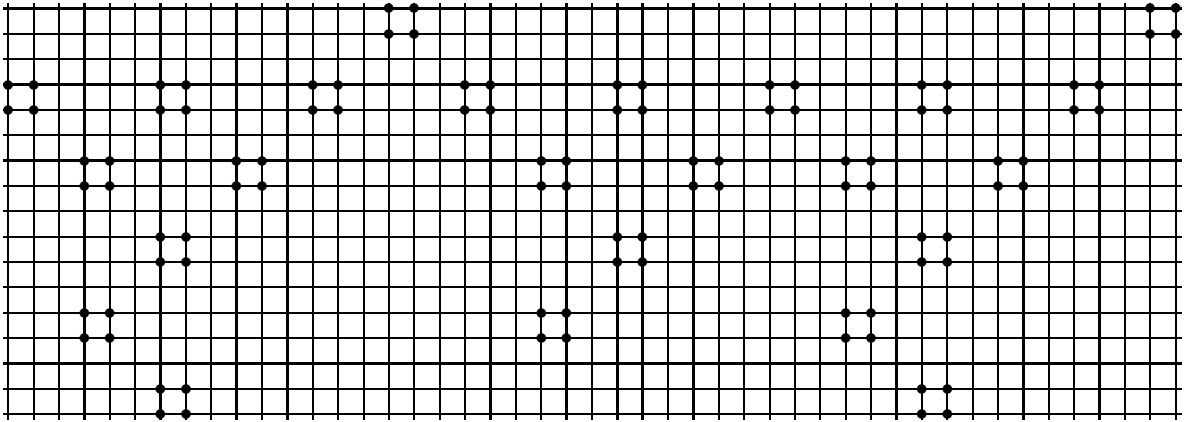
\beginpicture(46.5,16.5)(-.25,-.25) \thinlines
\multiput(-.2,0)(0,1){17}{\line(1,0){46.4}}
\multiput(0,-.2)(1,0){47}{\line(0,1){16.4}}
\\(0,12) \\(3,3) \\(3,9) \\(6,0) \\(6,6) \\(6,12) \\(9,9) \\(12,12) \\(15,15)
\\(18,12) \\(21,3) \\(21,9) \\(24,6) \\(24,12) \\(27,9) \\(30,12) \\(33,3)
\\(33,9) \\(36,0) \\(36,6) \\(36,12) \\(39,9) \\(42,12) \\(45,15)
\endpicture
$$
in which there are only two solutions, ``positive'' and ``negative.''
This construction provides a way to represent the values of boolean
variables in a satisfiability problem.

We can now use Lichtenstein's theorem that planar {\tt 3SAT} 
is NP-complete~[6].
An instance of planar {\tt 3SAT} is a set of variables 
$v1,\ldots,vn$ arranged
in a straight line, together with a set of three-legged clauses above and below
them, where the clauses are properly nested so that none of the legs
between clauses and variables cross each other. We can always put the clauses into
a rectilinear configuration such as
$$\unitlength=6pt
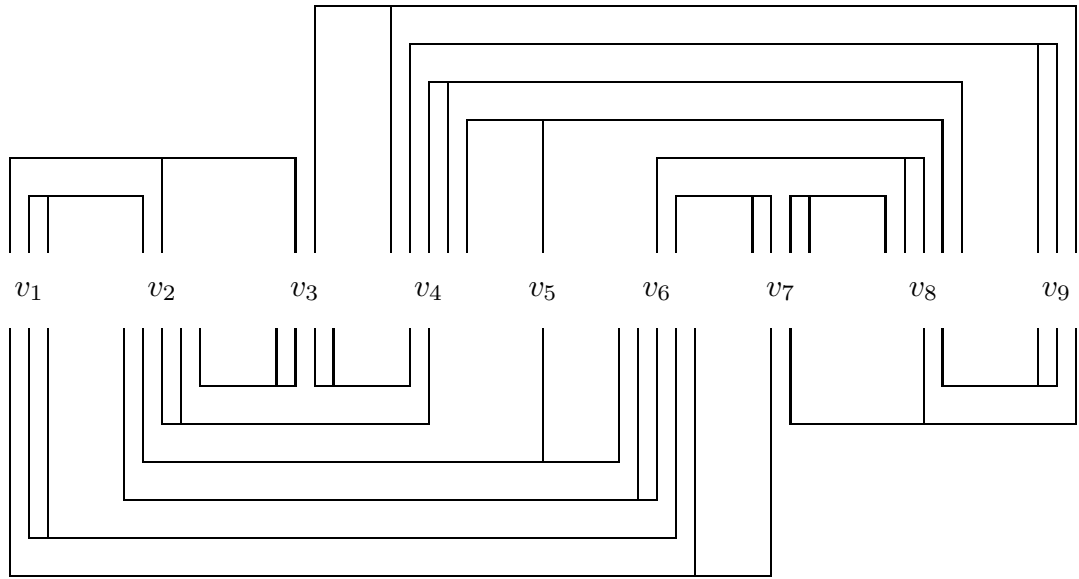
\beginpicture(58,34)(0,-17)
\def\\(#1,#2,#3:#4){\ii=#4 \advance\ii-2
 \put(#1,2){\line(0,1){\number\ii}}
 \put(#2,2){\line(0,1){\number\ii}}
 \put(#3,2){\line(0,1){\number\ii}}
 \jj=58 \advance\jj-#1 \put(\number\jj,-2){\line(0,-1){\number\ii}}
 \jj=58 \advance\jj-#2 \put(\number\jj,-2){\line(0,-1){\number\ii}}
 \jj=58 \advance\jj-#3 \put(\number\jj,-2){\line(0,-1){\number\ii}}
 \ii=#3 \advance\ii-#1 \put(#1,#4){\line(1,0){\number\ii}}
 \put(\number\jj,-#4){\line(1,0){\number\ii}}}%
\\(1,9,16:7)
\\(2,3,8:5)
\\(17,21,57:15)
\\(22,55,56:13)
\\(23,24,51:11)
\\(25,29,50:9)
\\(35,48,49:7)
\\(36,40,41:5)
\\(42,43,47:5)
\def\\(#1:#2){\put(#1,0){\makebox(0,0){$v_{#2}$}}}%
\\(2:1) \\(9:2) \\(16.5:3) \\(23:4) \\(29:5)
\\(56:9) \\(49:8) \\(41.5:7) \\(35:6)
\endpicture
$$
which corresponds to Lichtenstein's ``crossover box'' [6, Figure~5].

We construct an instance of {\tt MFL} from a given instance of planar 
{\tt 3SAT}
by representing the vertical legs for each variable as chains of four-point
clusters; this guarantees that each variable will have one of two values,
corresponding to the common orientation of all its clusters. We can easily
stretch out the diagram so that there is no interference between the variables
except at places where three legs of a clause come together in a horizontal
segment.

It remains to specify the representation of the clauses. By symmetry we need
only describe the representation that appears above the variables. Each
horizontal section of a comb-like clause in the upper portion will be
represented by a configuration of the form
$$
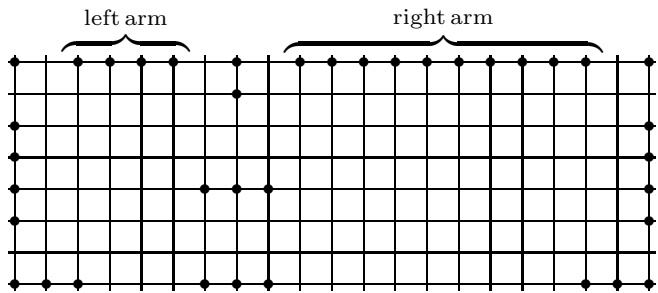
\beginpicture(20.5,8.5)(-.25,-.25)
\multiput(-.2,0)(0,1){8}{\line(1,0){20.4}}
\multiput(0,-.2)(1,0){21}{\line(0,1){7.4}}
\def\\(#1,#2){\put(#1,#2){\dsk}}%
\\(0,0) \\(1,0) \\(2,0) \\(6,0) \\(7,0) \\(8,0) \\(18,0) \\(19,0) \\(20,0)
\\(0,2) \\(0,3) \\(0,4) \\(0,5) \\(20,2) \\(20,3) \\(20,4) \\(20,5)
\\(6,3) \\(7,3) \\(8,3) \\(7,6)
\\(0,7) \\(2,7) \\(3,7) \\(4,7) \\(5,7) \\(7,7)
\\(9,7) \\(10,7) \\(11,7) \\(12,7) \\(13,7) \\(14,7) \\(15,7)
\\(16,7) \\(17,7) \\(18,7) \\(20,7)
\put(3.5,8){\makebox(0,0){$\overbrace{\hskip4\unitlength}^{\rm left\,arm}$}}
\put(13.5,8){\makebox(0,0){$\overbrace{\hskip10\unitlength}^{\rm right\,arm}$}}
\endpicture
$$
with $6l+4$ dots in the left arm and $6m+4$ dots in the right arm, for some~$l$
and~$m$. (The 
three triples at the bottom will connect to clusters that represent
variables, as explained below. Those clusters  will occur at positions
that are congruent mod~6; the arms of a comb 
stretch out so that they reach the variables appropriate to the clause.)

In each group of three dots at the bottom of this construction, the arrow
for the middle dot must go either up or down. All three middle arrows
cannot go up, because that forces
$$
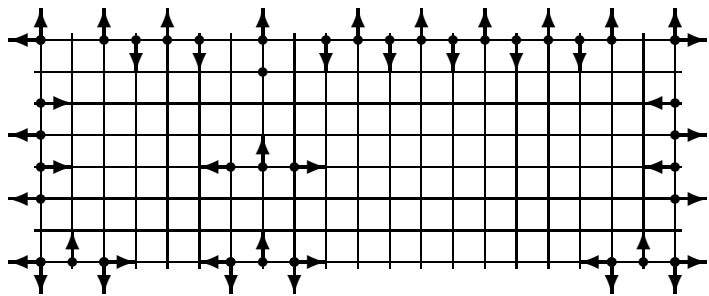
\beginpicture(22.5,9.5)(-1.25,-1.25)
\multiput(-.2,0)(0,1){8}{\line(1,0){20.4}}
\multiput(0,-.2)(1,0){21}{\line(0,1){7.4}}
\thicklines
\lfvec(0,0) \dnvec(0,0) \upvec(1,0) \dnvec(2,0) \rtvec(2,0)
 \lfvec(6,0) \dnvec(6,0) \upvec(7,0) \dnvec(8,0) \rtvec(8,0)
 \lfvec(18,0) \dnvec(18,0) \upvec(19,0) \dnvec(20,0) \rtvec(20,0)
\lfvec(0,2) \rtvec(0,3) \lfvec(0,4) \rtvec(0,5)
 \rtvec(20,2) \lfvec(20,3) \rtvec(20,4) \lfvec(20,5)
\lfvec(6,3) \upvec(7,3) \rtvec(8,3) \put(7,6){\dsk}
\lfvec(0,7) \upvec(0,7) \upvec(2,7) \dnvec(3,7) \upvec(4,7) \dnvec(5,7)
 \upvec(7,7) \dnvec(9,7) \upvec(10,7) \dnvec(11,7) \upvec(12,7) \dnvec(13,7)
 \upvec(14,7) \dnvec(15,7) \upvec(16,7) \dnvec(17,7) \upvec(18,7)
 \upvec(20,7) \rtvec(20,7)
\endpicture
$$
and there is no way to attach an arrow to the middle dot in the second row.

However, there are solutions in which any one of the middle arrows goes down.
For example, we can choose
$$
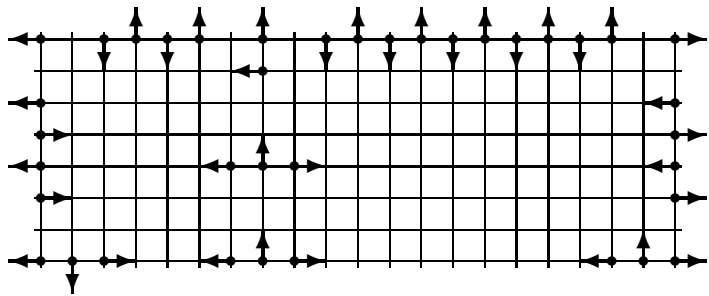
\beginpicture(22.5,9.5)(-1.25,-1.25)
\multiput(-.2,0)(0,1){8}{\line(1,0){20.4}}
\multiput(0,-.2)(1,0){21}{\line(0,1){7.4}}
\thicklines
\lfvec(0,0) \dnvec(1,0) \rtvec(2,0)
 \lfvec(6,0) \upvec(7,0) \rtvec(8,0)
 \lfvec(18,0) \upvec(19,0) \rtvec(20,0)
\rtvec(0,2) \lfvec(0,3) \rtvec(0,4) \lfvec(0,5)
 \rtvec(20,2) \lfvec(20,3) \rtvec(20,4) \lfvec(20,5)
\lfvec(6,3) \upvec(7,3) \rtvec(8,3) \lfvec(7,6)
\lfvec(0,7) \dnvec(2,7) \upvec(3,7) \dnvec(4,7) \upvec(5,7)
 \upvec(7,7) \dnvec(9,7) \upvec(10,7) \dnvec(11,7) \upvec(12,7) \dnvec(13,7)
 \upvec(14,7) \dnvec(15,7) \upvec(16,7) \dnvec(17,7) \upvec(18,7)
 \rtvec(20,7)
\endpicture
\advance\belowdisplayskip-9pt
$$
or 
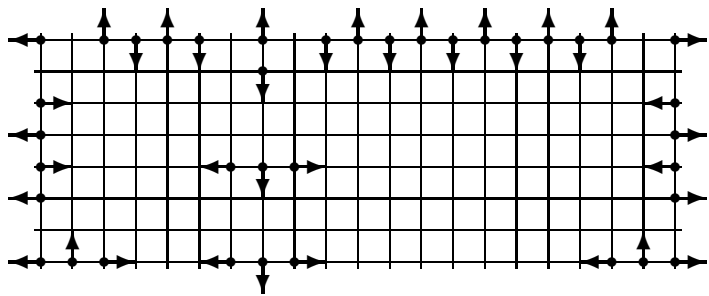
$$\beginpicture(22.5,9.5)(-1.25,-1.25)
\multiput(-.2,0)(0,1){8}{\line(1,0){20.4}}
\multiput(0,-.2)(1,0){21}{\line(0,1){7.4}}
\thicklines
\lfvec(0,0) \upvec(1,0) \rtvec(2,0)
 \lfvec(6,0) \dnvec(7,0) \rtvec(8,0)
 \lfvec(18,0) \upvec(19,0) \rtvec(20,0)
\lfvec(0,2) \rtvec(0,3) \lfvec(0,4) \rtvec(0,5)
 \rtvec(20,2) \lfvec(20,3) \rtvec(20,4) \lfvec(20,5)
\lfvec(6,3) \dnvec(7,3) \rtvec(8,3) \dnvec(7,6)
\lfvec(0,7) \upvec(2,7) \dnvec(3,7) \upvec(4,7) \dnvec(5,7)
 \upvec(7,7) \dnvec(9,7) \upvec(10,7) \dnvec(11,7) \upvec(12,7) \dnvec(13,7)
 \upvec(14,7) \dnvec(15,7) \upvec(16,7) \dnvec(17,7) \upvec(18,7)
 \rtvec(20,7)
\endpicture
$$
and there is a third solution that is essentially a mirror image of the first.

We can place four-point clusters below a row of three dots in such a way that
a downward arrow on the top middle dot forces an orientation on the clusters, but an
upward arrow on the top middle dot forces nothing:
$$
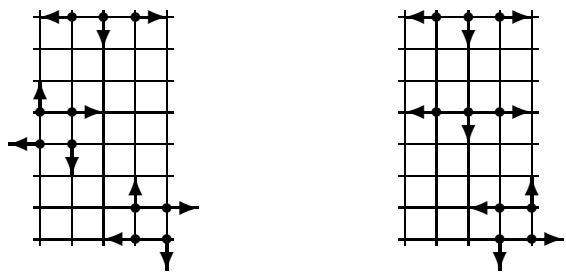
\beginpicture(6.5,8.5)(-1.25,-1.25) \thinlines
\multiput(-.2,0)(0,1){8}{\line(1,0){4.4}}
\multiput(0,-.2)(1,0){5}{\line(0,1){7.4}}
\thicklines
\lfvec(1,7) \dnvec(2,7) \rtvec(3,7)
\lfvec(0,3) \upvec(0,4) \rtvec(1,4) \dnvec(1,3)
\lfvec(3,0) \upvec(3,1) \rtvec(4,1) \dnvec(4,0)
\endpicture
\hskip 5\unitlength
\beginpicture(6.5,8.5)(-1.25,-1.25) \thinlines
\multiput(-.2,0)(0,1){8}{\line(1,0){4.4}}
\multiput(0,-.2)(1,0){5}{\line(0,1){7.4}}
\thicklines
\lfvec(1,7) \dnvec(2,7) \rtvec(3,7)
\lfvec(1,4) \dnvec(2,4) \rtvec(3,4)
\lfvec(3,1) \upvec(4,1) \rtvec(4,0) \dnvec(3,0)
\endpicture
$$
By choosing one of these junction configurations for each variable in the
clause, depending on whether the variable is negated or not, we obtain an
instance of {\tt MFL} that has a solution if and only if the given planar
clauses are satisfiable. 

\bigskip\noindent
{\bf Backtracking.}\enspace
We have now proved that {\tt MFL} is NP-hard. However, in practice
a solution or proof of nonexistence can often be found quickly by backtracking,
using the idea of ``preclusion'' introduced by Golomb and Baumert~[1].
When a trial value $xj$ is selected from~$Aj$, it precludes all selections
of other~$xk$ that are incompatible with it; precluded values can be 
(temporarily) removed from~$Ak$. The problem of compatible
representatives is precisely the abstract general setting that
supports this notion of preclusion.

Golomb and Baumert suggest choosing $xj$ at each stage from a currently
smallest set~$Aj$ whose representative has not yet been chosen. If we are
simply looking for a solution, 
instead of
enumerating all solutions, it would also be worthwhile to select elements
that preclude as few others as possible.

For example, if an element of $Aj$ doesn't preclude any others, we can set
$xj$ equal to that element without loss of generality. If 
$x\in Aj$ precludes
only one element $y\in Ak$  and no others,
and if we find no solution when $xj=x$,
then we can set $xk=y$ without loss of generality.

\bigskip\noindent
{\bf Further work.}\enspace
A recent paper by Simon [8] considers the assignment of channels to 
transmitters
in a radio communication system. This is another case of a 
compatibility problem,
rather like the mapmaker's problem because nearby transmitters must not
broadcast on the same channel. Simon presents a polynomial-time approximation
scheme that is  guaranteed to find at least a fixed fraction of the optimum
number of compatible channels. This suggests that many useful approximation
schemes for other instances of the general compatibility problem might
 remain to be found.

\bigskip
\centerline{\bf References}

\smallskip
\disleft 20pt:
[1]:{\sc Solomon Golomb {\rm and} Leonard D. Baumert}, 
{\sl  Backtrack programming},
Journal of the ACM, 12 (1965), pp.~516--524.

\smallskip
\disleft 20pt:
[2]:{\sc Martin Gr\"otschel, L\'aszl\'o Lov\'asz, {\rm and} 
Alexander Schrijver},
{\sl Geometric Algorithms and Combinatorial Optimization}, Springer-Verlag,
New York, 1987.

\smallskip
\disleft 20pt:
[3]:{\sc Marshall Hall, Jr}, {\sl Combinatorial Theory}, Blaisdell,
1967. Second edition, Wiley-Interscience, 1986.

\smallskip
\disleft 20pt:
[4]:{\sc P. Hall}, {\sl On representatives of subsets}, Journal of the
London Mathematical Society, 10 (1935), pp.\ 26--30.

\smallskip
\disleft 20pt:
[5]:{\sc Donald E. Knuth},
{\MFbook\/},
Addison-Wesley, 
Reading, Mass.,
1986.

\smallskip
\disleft 20pt:
[6]:{\sc David Lichtenstein}, 
{\sl Planar formul{\ae} and their uses}, SIAM
Journal on Computing,      11  (1982), pp.~329--343.

\smallskip
\disleft 20pt:
[7]:{\sc G. Minty}, {\sl  On maximal independent sets of vertices in claw-free
graphs}, Journal of Combinatorial Theory    (B),     28  (1980),
pp.~284--304.

\smallskip
\disleft 20pt:
[8]:{\sc Hans Ulrich Simon}, {\sl Approximation algorithms for channel
assignment in cellular radio networks}, in Fundamentals of Computation
Theory, Proceedings of FCT'89, edited by J.~Csirik, J.~Demetrovics,
and F.~G\'ecseg, Lecture Notes in Computer Science 380 (1989), pp.\
405--415.

\bye

\magnification=\magstep1

\parskip 2pt

\noindent
{\bf Connection to satisfiability.}\enspace
We have observed that the problem of compatible representatives with
each $\|Aj\|2$ reduces to $2\;$SAT. In general, if each $\|Aj\|k$ and $k2$,
the problem reduces directly to an instance of $k\;$SAT in which each literal
occurs positively just once. The literals are $(j,a)$ for
$a\in Aj$, and the clauses are
$$\vcenter{\halign{\hfil${\displaystyle#}$\qquad&#\hfil\cr
\bigvee{a\in Aj}(j,a)\,,&for $1jn\,;$\cr
\noalign{\medskip}
\overline{(j,a)}\vee\overline{(k,b)}\,,&for $1j<kn$ 
and $a$ incompatible with~$b$.\cr}}$$ 

Conversely, any instance of $k\;$SAT with $m$ clauses reduces to the
compatibility problem of finding representatives
$(x1,\ldots,xm)$, with $xj$ a number
of the $j$th clause and with two literals compatible iff they aren't
negatives of each other.

\bye